\documentclass[prb,twocolumn,showpacs]{revtex4-1}
\usepackage{amsmath}
\usepackage[dvips,xdvi]{graphicx}
\usepackage{amssymb}
\usepackage{epsfig}

\begin{document}
\title{Edge states and flat bands in graphene nanoribbons with arbitrary geometries}

\author{W. Jaskolski$^1$}
\email{wj@fizyka.umk.pl}
\author{A. Ayuela$^{2}$}
\author{M. Pelc$^1$}
\author{H. Santos$^3$}
\author{L. Chico$^3$}

\affiliation
{
$^1$ Instytut Fizyki UMK, Grudzi\c{a}dzka 5, 87-100 Toru\'n, Poland\\
$^2$ Centro de F\'\i sica de Materiales CFM-CPM CSIC-UPV/EHU, Departamento de F\'\i sica de Materiales(Facultad de  Qu\'\i mica, UPV) and Donostia International Physics Center, 20080 San Sebastian/Donostia, Spain\\
$^3$ Instituto de Ciencias de Materiales de Madrid, CSIC, Cantoblanco, 28049 Madrid, Spain
}
\date{\today}

\begin{abstract}

We prescribe general rules to predict the existence of  edge states and zero-energy flat bands in graphene nanoribbons 
and graphene 
edges of arbitrary shape. No calculations are needed. For 
the so-called {\it{minimal}} edges, 
the projection of the edge translation vector into the zigzag direction of graphene
uniquely determines the edge bands. 
By adding extra nodes to minimal edges, arbitrary modified edges can be obtained. 
The edge bands of modified graphene edges can be found by applying hybridization rules of the extra atoms with the ones belonging to the original edge.
Our prescription correctly predicts the localization 
and degeneracy 
of the zero-energy bands at one of the graphene sublattices, confirmed by 
tight-binding and first-principle calculations. It also allows us to qualitatively predict the existence of $E\ne 0$ bands 
appearing in the energy gap of certain 
edges and 
nanoribbons.

\end{abstract}

\maketitle

\section {Introduction}

Graphene is presently one of the most studied materials in condensed matter and materials science. It presents a plethora of interesting physical phenomena 
due to the fact that its elementary electronic excitations behave as two-dimensional chiral Dirac fermions. \cite{transport}
Graphene nanoribbons (GNR),  
stripes of nanometric widths cut from graphene, are also 
the subject of a growing interest.
They exhibit edge-localized states, 
which may have potential practical applications and are the key ingredient in many of the fascinating properties of graphene and its nanometric derivatives. Edge states play an important role in transport and magnetic properties
%WJ_1:  "of GNRs" - added instead of "their"
of GNRs,
such as the quantum Hall effect \cite{qhe} and the quantum spin Hall effect. \cite{qshe} The magnetic properties of nanoribbons are directly related to the existence of localized edge states. \cite{magnetic}

The appearance of the edge states in 
GNR have been investigated long before \cite{fujita_1996,nakada} graphene sheets 
were 
experimentally achieved.  \cite{novoselov} A large number of theoretical works 
based on 
continuum Dirac-like, \cite{brey_2006} tight-binding, \cite{tb_zheng_2007} and density-functional \cite{son_2006} approaches have been applied to study GNRs with different edges, such as zigzag, \cite{brey_2006,son_2006} armchair,\cite{brey_2006,son_2006,tb_zheng_2007} or mixed with cove and with Klein nodes. \cite{cresti_2009} 
All these edge terminations have been experimentally identified by different techniques, such as scanning tunneling microscopy, \cite{niimi_2006,kobayashi_2006}    
high resolution transmission electron microscopy \cite{liu_2009}  or atom-by-atom spectroscopy. \cite{suenaga_2010}

From the theoretical viewpoint, it is important to identify general edges and nanoribbons which present localized edge states, as well as their degeneracy and characteristics. Although the boundary conditions for an important subset of edge terminations has been studied, \cite{akhmerov} as well as certain 
modifications  \cite{wakabayashi_2010} with experimental interest, \cite{liu_2009} 
%WJ_2: - order of words changed
until now identifying general ribbons with edge states and their band structure characterization is still an open question.

In this paper we solve this problem by giving  
a simple prescription which allows to predict the existence of the edge states 
and their degeneracies 
in a given
 graphene edge or nanoribbon. We show that no calculations are needed to find out whether the edge states and flat bands exist at the Fermi energy ($E_F$) for any kind of periodic graphene 
edge or 
%WJ_3: added
wide enough 
nanoribbon with noninteracting edges, at least at the level of $\pi$-electron approximation. 

We consider periodic edges defined by a translation vector {\bf T}. 
Our approach follows two steps: 
First, we 
characterize 
{\it {minimal}} edges, \cite{akhmerov}
i.e., those with a minimum number of edge nodes and dangling bonds per translation period. 
For minimal edges, the spectrum of $E=0$ flat bands is 
determined by the zigzag edge component of ${\bf T}$, i.e., the projection of ${\bf T}$ along the zigzag direction,  which poses a 
folding rule to the graphene zigzag edge band.
Next, we show that any other edge can be obtained from a minimal one by adding extra nodes. We call these {\it{modified}} edges. The extra nodes provide extra bands at the Fermi energy that may hybridize with the edge orbitals. 
If hybridization takes place, the extra bands couple with the existing $E=0$ edge bands and split in energy, moving towards the bulk bands.
Such splitting depends on whether the extra nodes belong to the same sublattice as 
that where the edge zero-energy bands are localized.
We find an extremely simple rule 
to determine, without performing any calculations, 
the existence, origin and localization properties of edge states and flat bands of any modified edges, thus allowing the complete characterization
of the low-energy properties of any GNR with general edges.

This prescription for the
identification of 
edge states and flat bands in graphene edges and nanoribbons was found after performing calculations for a large number of different GNRs. The calculations were performed in the tight-binding (TB) $\pi $-electron approximations using hopping parameter $t_0=2.66$ eV.  We have collected a huge amount of data, but only some of them are presented here for illustration purposes.

The rest of the paper is organized as follows. In Sec. \ref{sec:fold} we give the geometrical description of the edges and nanoribbons as well as the 
%WJ_4:  order of words changed
edge band 
folding rule with an example regarding a minimal edge. Sec. \ref{sec:mod} describes modified (i.e., non-minimal) edges, starting with zigzag edges with Klein defects and cape structures, which allows us to set the rules 
%WJ_5 change to remove two close "to"
%LC i am sorry but gramatically it has to be a to and there is no problem with that
%for finding
to find 
out the zero-energy edge bands.
%and their degeneracy. 
%WJ_6: a sentende about the diagrams added
%LC typo corrected 
%degenaracy
We introduce simple diagrams, which determine their localization and degeneracy.
The Section concludes discussing edge bands away from the Fermi energy, focusing on armchair modified and chiral edges. Finally, In Sec. \ref{sec:sum} we summarize our results.

\section{Characterization of 
graphene
edges. Folding rule and minimal edges}
\label{sec:fold} 

\begin{figure}[htbp]
\includegraphics*[width= 48mm,angle=-90]{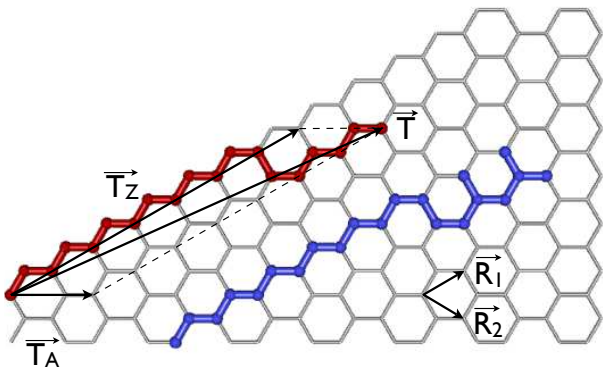}
\caption{({\it{Color online}})
Graphene lattice and the translation vector {\bf T} 
$(n,m)=(8,1)$. The vectors {\bf{T}$_Z$} and {\bf{T}$_A$} are projections of {\bf T} into the zigzag and armchair directions. The primitive lattice vectors {\bf{R}$_1$} and {\bf{R}$_1$} are also shown so that {\bf T}=
$n${\bf R}$_1+m${\bf R}$_2$. The 
dark-grey 
(red) line shows the {\it{minimal}} edge; the 
light-grey
(blue) line shows a modified edge constructed by attaching two Klein nodes to the minimal case. } \label{fig1}
\end{figure}

A graphene edge consists of
a set of lattice sites with only one or two neighbors, i.e., having two or one dangling bonds, respectively. 
In this work we assume that the edge atoms are arranged periodically. Therefore, they form a one-dimensional superlattice with a translation period defined by ${\bf T}=n{\bf R}_1+m{\bf R}_2$, where ${\bf R}_1$ and ${\bf R}_2$ are the primitive vectors of the honeycomb lattice, as seen in Fig. \ref{fig1}. 
For a given period ${\bf T}$ with indices $(n,m)$, the number of edge sites $N_e$ and the number of dangling bonds $N_d$ can be arbitrarily large, but neither of them can be smaller than $n+m$.  \cite{akhmerov} Following Ref. \onlinecite{akhmerov}, when the number of edge atoms equals that of dangling bonds
{\em and} both are equal to $n+m$ 
($N_e=N_d=n+m$), we call the edge {\it{minimal}}. For a minimal edge, the number of nodes in one sublattice equals $n$ and the number of nodes in the other sublattice equals $m$.

Any minimal edge can be modified by adding extra edge nodes. To identify 
a modified edge one has just to add information on the extra nodes added to the corresponding minimal edge.
As an example, Fig. \ref{fig1} presents two edges associated to the translation vector ${\bf T}$ $(8,1)$. The minimal edge is marked by a red line. Another possible edge, with two additional 
nodes, constituting the so-called Klein defects,
is marked by a blue line.

The translation vector ${\bf T}=n{\bf R}_1+m{\bf R}_2$ defining any graphene edge can be decomposed into two 
important 
directions in the honeycomb lattice
(see Fig. \ref{fig1}), 
namely the armchair and zigzag: ${\bf T} = {\bf T}_A + {\bf T}_Z$. Note that ${\bf T}_Z = (n-m) {\bf R}_1$ and 
${\bf T}_A = m ({\bf R}_1 + {\bf R}_2)$. 
This decomposition is crucial in our analysis, since
 it is well-known that an armchair edge does not have $E=0$ localized states, while the zigzag termination reveals a flat edge band at Fermi energy (in 
  %LC article the added
 the 
 $\pi $ electron approximation) for the wavevector $k>2/3 \pi$, as shown in 
previous nanoribbon band structure calculations \cite{fujita_1996,nakada}. 
Also, it is easy to see that a minimal edge corresponding to 
${\bf T}$  is a simple combination of the zigzag edge along 
${\bf T}_Z$ and the armchair edge along  ${\bf T}_A$. 
%Decomposition of ${\bf T}$ into ${\bf T}_A$ and ${\bf T}_Z$ 
%helps to understand the concept of minimal edge, which 
%can be seen as a simple combination of the zigzag edge 
%along ${\bf T}_Z$ and the armchair edge along  ${\bf T}_A$. 

\begin{figure}[htbp]
\includegraphics*[width= 80mm]{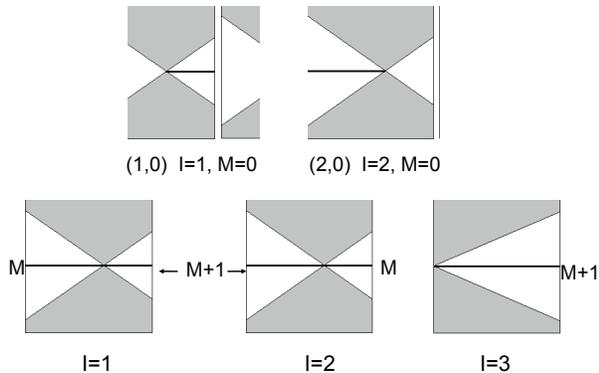}
\caption{Schematic band structures of 
zigzag (1,0), (2,0) and general 
$(S,0)$ 
edges 
after folding 
the (1,0) zigzag 
edge band,
where $S=I+3M$ as described in the text. The shaded areas represent the band continuum of states. 
Degeneracies of the zero-energy bands ($M$, $M+1$) are indicated in the lower panels close to the corresponding edge bands at the Fermi energy.
They correspond to semi-infinite graphene sheet with only one edge. In case of a GNR with equal edges the degeneracies are doubled.}\label{rule}
\end{figure}

The 
%WJ_7: two words added
schematic
spectrum 
(close to $E_F$)
of the zigzag
edge defined by the smallest ${\bf T}_Z =  {\bf R}_1$, 
i.e., the edge (1,0),  
is shown in the upper left panel of Fig. \ref{rule}. 
The spectrum of a zigzag 
$(n-m,0)=(S,0)$ 
edge  
defined by ${\bf T}_Z =  S{\bf R}_1$ is obtained by folding this spectrum $S$-times. 
The (2,0) case is shown explicitly in the upper right panel of Fig. \ref{rule}. Both, the (1,0) and the (2,0) edge, have degeneracy 1.
By repeated folding of the minimum (1,0) zigzag edge one can easily find the bandstructure and edge band degeneracies of a general 
$(S,0)$ 
edge.   
Any integer number 
$S>0$ can be written as $S=I+3M$, where $I=1,2,3$ and $M=0,1,2,...$.
For $I=1$ and 2 the folded spectrum has always the Dirac point at $2/3 \pi$ while for $I=3$ the Dirac point is at $k=0$, as illustrated in Fig. \ref{rule}. The schematic band structures of all zigzag edges 
obtained
from this folding are shown in Fig. \ref{rule}. The degeneracies of the zero-energy band are also indicated in the lower panels 
%WJ_8: math-mode added
($M$ and $M+1$) close to the corresponding edge bands.

Since
the armchair component does not provide any edge states, one can expect that the spectrum of a minimal-edge $(n,m)$ 
will be similar to the spectrum of the $(n-m,0)$ zigzag 
edge 
at least close to $E_F$. 
We have performed tight-binding calculations for a large number of different minimal-edge GNRs,
and verified
that the presented above folding rule holds in all the cases considered. 
Notice that for a graphene nanoribbon with two equal edges, the degeneracy of the flat band is twice that of the isolated edge, provided that the 
ribbon is wide enough to neglect interaction between the edges. 

\begin{figure}[htbp]
\includegraphics*[width= 60mm,angle=90]{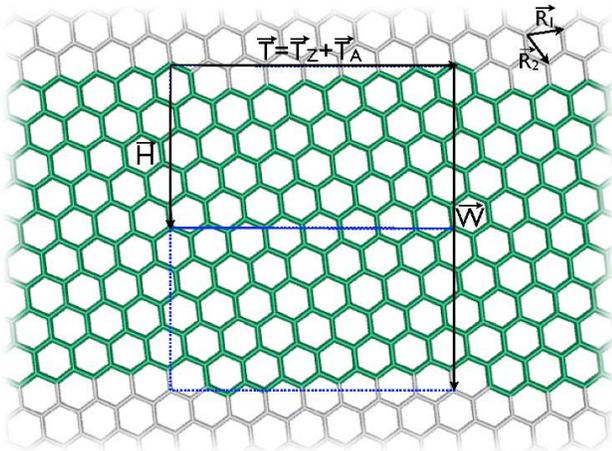}
\caption{({\it{Color online}}) Geometric structure of the  2(7,1) GNR highlighted in dark grey (green) on a graphene sheet, indicating the translation vector ${\bf T}= {\bf T}_Z + {\bf T}_A$ and the width vector ${\bf W}=  2{\bf H}$, where  
%WJ_11: I think that decomposition of T is not neccessary
%${\bf H}=  -3{\bf R}_1 + 5{\bf R}_2$ 
${\bf H}$ 
is the smallest nanoribbon width vector belonging to the graphene lattice. 
%WJ_12: one sentence added, but may be it is not necessary
${\bf T}_Z = 6{\bf T}(1,0)$. 
The unit cells spanned by {\bf T} and {\bf H} or {\bf W} are marked.}
 \label{fig71}
\end{figure}

%WJ_9: new par. added
In order to construct ribbons corresponding to a particular edge ${\bf T}$, we first define the vector
${\bf H}$ as the smallest graphene lattice vector perpendicular to ${\bf T}$. The width of the ribbons studied here is spanned by a vector ${\bf W}$ given by an integer multiple of ${\bf H}$, ${\bf W}=N{\bf H}$
as it is shown in Fig. \ref{fig71}
%WJ_10: specification of the figure added
for the case of 2(7,1) GNR. 
For a fixed ${\bf T}$, ${\bf H}$ is uniquely determined up to a global 
plus or minus
sign; therefore, for our purposes, the ribbons with minimal edges are labeled by $N(n,m)$, where $N$ states the ribbon width, and $(n,m)$ indicates the minimal edge. 

Of course, there exist ribbons with minimal edge geometries, that may require a semi-integer $N$, 
such as the so-called antizigzag ribbons. 
As our main goal here is to study edge bands, we restrict ourselves to integer $N$, using values which yield non-interacting GNR edges.

\begin{figure}[htbp]
\includegraphics*[width= 80mm]{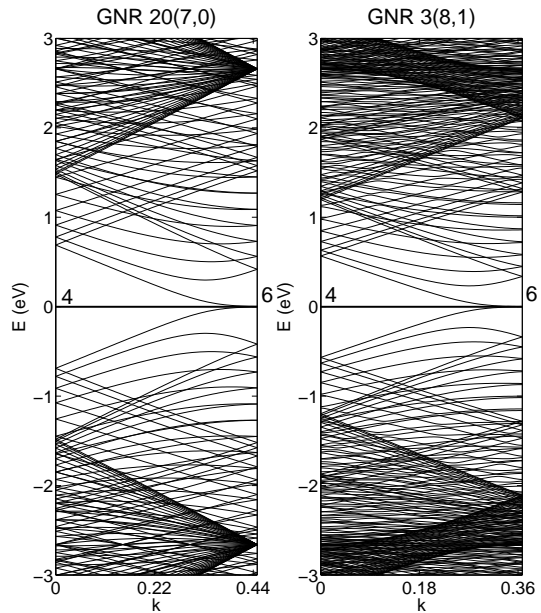}
%WJ_13: one word added "Energy"
\caption{Energy spectra of   
the 20(7,0) GNR (left) and 
the 3(8,1) GNR (right) close to the Fermi energy. The unit cells of both GNRs contain similar number of nodes. The degeneracies of the zero-energy edge bands (4 and 6) are indicated 
close to 
%WJ_14: I suggest Marta to change the figure and use k-values 
% as in
% the other figurs; also E -> Energy
% additionally, the 
% before that changes the fragment below must be changed
%$k=0$ and $k=\pi $.} \label{rib70_81}
extreme $k$ values.} \label{rib70_81}
\end{figure}

As a particular example, the spectrum of a nanoribbon with minimal-edge 3(8,1) is presented in the left panel of Fig. \ref{rib70_81}.  One can easily see that degeneracies of the flat $E=0$ bands are exactly the same as for the 20(7,0) zigzag GNR (right panel of Fig. \ref{rib70_81}). 
Also the gaps at $k=0$  and $\pi$ follow the folding rule to a large extent. 
In general, the minimal-edge $(n,m)$ GNR reveals the same zero-energy bands as the $(n-m,0)$ zigzag GNR, which in turn has the same spectrum as the $(n-m)$ times folded spectrum of the (1,0) zigzag nanoribbon of the same width.

\section{Modified edges}
\label{sec:mod}

\subsection{Coupling of edge defects and band splitting }

\begin{figure}[htbp]
\includegraphics*[width=80mm]{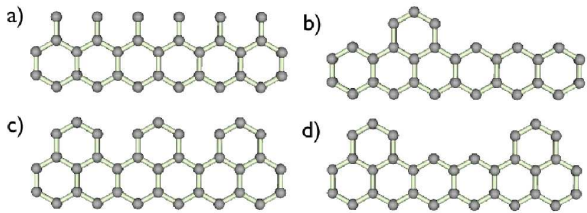}
\caption{Geometries of several modified zigzag graphene edges. (a) Bearded zigzag edge, composed of Klein defects; (b) a cape structure on a zigzag edge, obtained by bonding one extra atom to two adjacent Klein defects; (c) a cove edge; (d) a periodic modified edge with a cape.
} \label{fig:edges}
\end{figure}

Here we 
study a couple of modified zigzag edges. We start with the Klein defects,\cite{klein} which consist of atoms with only one neighbor, as depicted in 
Fig. \ref{fig:edges}(a). 
Then we proceed with other modified edges related to the so-called cove structures, which can be constructed by adding extra atoms to 
the Klein-edge zigzag nanoribbon. 
The basic unit of such modified edges is what we call a {\em cape}. 
This can be built by bonding 
two Klein defects
to one extra atom, 
and the resulting structure is depicted 
in Fig. \ref{fig:edges}(b). When extra atoms are added every two Klein defects, one gets the 
cove edge, shown in Fig. \ref{fig:edges}(c). We will consider modified edges where capes are more separated than in the cove edge structure, as in Fig. \ref{fig:edges}(d). 
Based on these examples we discuss how such modifications influence the mixing and splitting of states at $E=0$. In order to perform numerical 
calculations, we choose wide ribbons with equal edges. As discussed in the previous Section, if the ribbons are wide enough, the interaction between edges is 
reduced; and as they have equal edges, the degeneracies are obviously twice those stated in Sec. \ref{sec:fold}. 
 
\subsubsection {Zigzag edge with Klein defects} 

\begin{figure}[htbp]
\includegraphics*[width= 80mm]{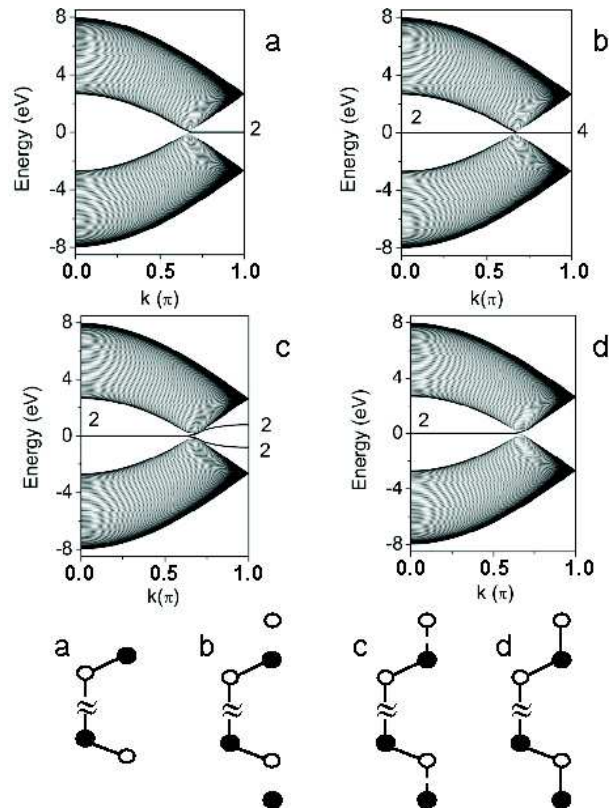}
\caption{Evolution of the spectrum of 40(1,0) zigzag GNR (a) to the spectrum of 40(1,0) GNR with Klein nodes attached at both sides (d). In (b), two extra non-attached nodes introduce a doubly degenerate $E=0$ flat band. In (c), connecting the extra nodes with $t=0.5t_0$ couples and split the flat bands in the range of $k>2/3 \pi $. The bottom panel illustrates how the extra nodes are attached to the upper and lower part of the unit cell of the zigzag GNR.} \label{fig3}
\end{figure}

We consider first a simple zigzag nanoribbon modified by Klein defects.
The spectrum of a zigzag nanoribbon of width 40 and minimal edge, i.e., a 40(1,0) GNR, is presented in Fig. \ref{fig3}(a). The two zero-energy edge bands localized at opposite edges\cite{nakada} of the GNR extend for $k>2/3 \pi$. The condition to get $E=0$ requires a non-vanishing amplitude of the wavefunction in one of the two graphene sublattices\cite{wakabayashi_2001}. For $k=\pi$ the corresponding wavefunctions are localized at the edge nodes. 
Opposite zigzag edges have atoms belonging to different sublattices.
For $k$ closer to $2/3 \pi$ the wavefunctions 
corresponding to these edge bands 
penetrate more into the 
inner part of the GNR;
therefore, the edge states interact more for these lower $k$ values, 
mixing and splitting 
into bonding and antibonding bands.

When Klein defects are added to both sides of the GNR, the flat bands appear from $k=0$ to $2/3 \pi$. In order to understand 
this change on the flat $E=0$ bands with respect to $k$ values, we gradually modify the coupling of the extra atoms conforming the Klein bearded edge. 

We first add a Klein node at each side of the GNR unit cell, 
but setting the hoppings equal to zero. As the on-site energies of these extra atoms are set to zero, 
an additional doubly degenerated zero-energy band appears,  as  shown 
 in the spectrum of Fig. \ref{fig3}(b). The double degeneracy is due to the contribution of the two edges. 
Let us now 
switch on the hopping, setting 
$t=0.5 t_0$. The resulting spectrum is shown in Fig. \ref{fig3}(c).  
The hopping connects atoms 
that belong to different sublattices; 
this allows for the interaction of
the corresponding flat bands, 
which 
hybridize and split. 
Such splitting is more pronounced for $k=\pi$, since their localization and overlap is stronger than for any other $k$, and increases gradually from 
 $k=2/3 \pi$ to the edge of the Brillouin zone. 
Finally, when $t=t_0$, see Fig. \ref{fig3}d, the splitting is so strong that the %aay interacting 
bonding and antibonding bands 
interact
from $k=2/3 \pi$ to $k=\pi$
and
reach the continuum of bands. We end up with the spectrum of a zigzag GNR with Klein edges, which has a zero-energy flat band for $k< 2/3 \pi$.

%WJ_15: Leonor - analytical - ment TB, so I have added "TB" 
% it is analytical in the meaning of the TB hamiltonian matrix
% row
A closer inspection of the analytical TB solution for the $E=0$ band of zigzag GNR with Klein edges reveals that, contrary to the zigzag edge, the wavefunction never localizes only at the  Klein nodes. 
The wavefunction penetrates into GNR even for $k=0$: the damping factor equals 1/2 in this case and rises up to 1 for $k=2/3 \pi$. 

It is noteworthy  that if we modify only one edge 
by adding Klein nodes, the 
extra zero-energy band will be non-degenerate and will mix and split with only one of the edge bands
in the range of $k>2/3 \pi$. This mixed band is obviously 
localized at the edge where the Klein nodes are added.

\begin{figure}[htbp]
\includegraphics*[width= 80mm]{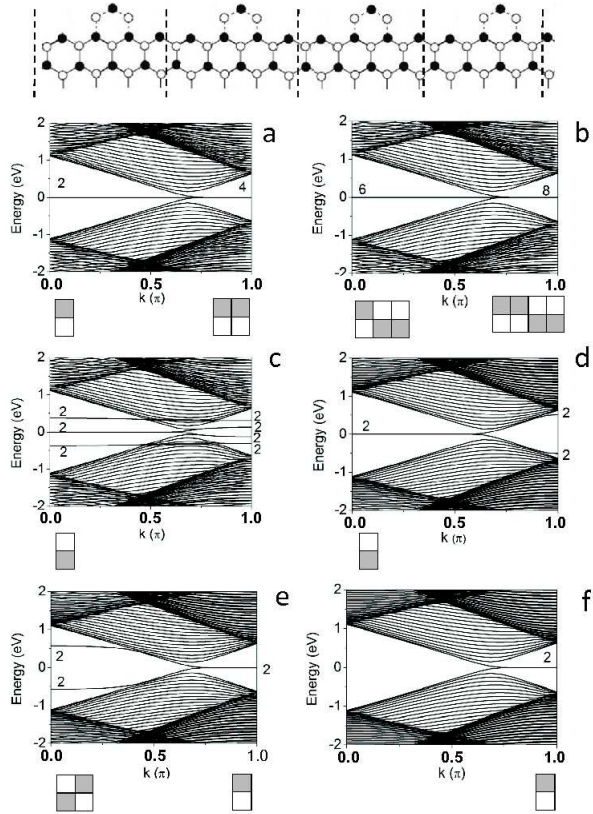}
\caption{Evolution of the spectrum of a (4,0) zigzag GNR when extra edge nodes are added to form a single 
cape structure. The upper panel shows the edge of GNR where connections to extra nodes are marked by dotted lines. The four unit cells are marked by dashed lines. The steps are the following: (a) pure (4,0) GNR, (b) with two disconnected extra nodes at each side of unit cell, (c) with two extra nodes 
(i.e. Klein defects) 
connected by $t=0.2 t_0$, (d) with two Klein nodes fully connected, (e) with another extra node connected to the Klein nodes by $t=0.2 t_0$  
forming the cape structure,
 and (f) with a cape structure added to the edge and all extra nodes fully connected. Diagrams 
%below (c) and (e) 
illustrate the mixing and splitting of flat bands, to be explained in the text.}\label{fig4}
\end{figure}

\subsubsection {Zigzag edge modified with a 
cape
structure}

%WJ_16: I have simplified the first sentence, but I am not sure
% if now is better 
%Although 
%we have already considered a modification of the zigzag ribbon 
%with Klein edges
%in the previous subsection, 
%we study here a more complex case. 
Here we consider a more complex modification of the zigzag edge.
It helps to illustrate and refine in detail the presented prescription, which allows us to predict 
the occurrence and degeneracies of 
edge states for any nanoribbon.  
The top panel of Fig. \ref{fig4} shows the construction of the cape at the upper edge only. In a similar way the cape is formed at the lower edge, but the black and open circels are reversed there.  

We have considered
the (4,0) zigzag GNR and a sequential addition of nodes to get the edges with a cape 
structure, as shown in Fig. \ref{fig4}. 
The procedure follows the  next sequence. 
One starts from a zigzag ribbon with quadrupled unit cell.
Two extra adjacent Klein nodes (open circles) are next added. 
Finally, the pair of Klein nodes is connected {\it{via}} 
another extra node (black circle), 
to form the cape structure. 

The spectrum around the Fermi level of the zigzag 40(4,0) GNR is presented in Fig. \ref{fig4}(a). 
It is obtained by folding four times the spectrum of Fig. \ref{fig3}(a).
The edge band degeneracies indicated in the Figure result from this folding. The two non-connected Klein nodes at each side of the GNR unit cell (a total of four extra nodes) 
 yields the addition of 
a four-fold degenerate 
flat band at zero energy. 
The degeneracies sum up to
six and eight at $k=0$ and $k=\pi$, respectively. This is shown in Fig. \ref{fig4}(b). 
We begin
by connecting these extra nodes %aay by 
with
a small hopping $t=0.2 t_0$ %aay. The bands are shown 
and plotting the bands 
in Fig. \ref{fig4}(c). For $k>2/3 \pi$  all the flat bands mix and split; 
they are still 
doubly 
degenerated since both edges of the GNR are equal. For $k< 2/3 \pi$ only two bands split and a doubly degenerate flat band survives at $E=0$. When $t=t_0$, see Fig. \ref{fig4}(d), all the split bands merge into the states continuum. 
The flat bands that survived at $E=0$
reveals that they are mainly localized at the extra Klein nodes and enter into GNR nodes of the same sublattice. Their  spreading
into the GNR is similar 
to that found in the flat bands 
of zigzag GNRs with Klein edges (see Fig. \ref{fig3}(d)).

To form the (4,0) GNR with a single 
cape 
structure we must add yet another extra node and connect it to the existing Klein nodes. %%%%% SUGGEST
Note that the extra node and the Klein nodes
belong to different sublattices.
A non-connected node adds a doubly degenerate zero-energy band to the spectrum of  Fig. \ref{fig4}(d). When connecting this extra atom to the previous Klein nodes with $t=0.5 t_0$,  the flat bands at $k< 2/3 \pi$,  
which are due to the Klein defects, 
hybridize with these %aay new 
added bands arising from the 
extra connected nodes
 and split.
This
splitting is shown in Fig. \ref{fig4}(e). For $t=t_0$, %aay see 
as seen in
Fig. \ref{fig4}(f).
Due to the strong coupling, the bands split 
so much that they merge into the continuum of states. 
This
doubly degenerate flat band at $k> 2/3 \pi$ 
 appears due to the introduction of the outermost extra atoms, 
 and as there were no localized 
 bands at that $k$ range which could mix with them, 
 they remain localized in the %aay new 
extra node, 
spreading into nodes 
 belonging to the 
 same sublattice.  

\begin{figure}[htbp]
\includegraphics*[width= 80mm]{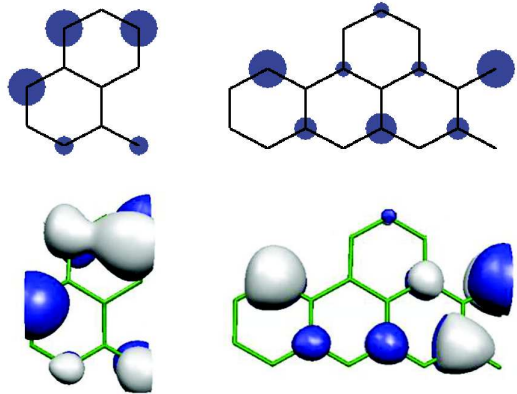}
\caption{({\it{Color online}})
Localization of the wavefunctions corresponding to the $E=0$ band at $k=\pi$ for 40(2,0) (left) and 40(4,0) (right) GNR with a cape
structure at the edges. 
%WJ_20B - the reference to Fig.5 added
The corresponding edges are shown in Fig. \ref{fig:edges} (c) and (d), respectively.
Only an edge and a few 
neighboring nodes in the GNRs unit cells are shown. 
%WJ_20C - the sentence below modified slightly
% and the order slightly changed
Upper panel: Results obtained using tight-binding method. Bottom panel: Results of first-principles calculations.
The dot diameter 
in the upper panel reflects the TB density at the nodes. 
No dot means that the wavefunction is exactly zero 
at this node.}
\label{fig8}
\end{figure}

This localization is confirmed by numerical calculations
performed within the tight-binding model, as well as using first-principles DFT approach\cite{dft}. 
The obtained wavefunctions
are shown in the right panel of  Fig. \ref{fig8}.
Their
localization in the 
(4,0) GNR with a single cape 
is different from the case of the cove edge,\cite{wakabayashi_2010} which can be 
considered as built from a (2,0) GNR 
%WJ_16B - reference to Fig.5d added
with a cape structure (see Fig. \ref{fig:edges}c). 
In Ref. \onlinecite{wakabayashi_2010} 
it was shown that
in the case of cove edge 
the wavefunctions of the zero-energy bands are not localized at the outermost 
edge atoms, 
 but at the neighboring nodes which belong to the 
 other 
 sublattice. 
We confirm this finding, both in the tight-binding and the DFT calculations 
shown in the left panel of Fig. \ref{fig8}
\cite{comment}.

%WJ_17: - new par.
One would naively expect
that the wave function should be more localized at the outermost atoms, which 
seems
more exposed to a chemical attack. However, it is easy to check that in a cove edge, the majority of the edge atoms are
not the outermost ones, but rather their nearest neighbors. 
The edge state is therefore localized
in the atoms closest to the 
outermost ones, which belong to the opposite sublattice, 
whereas the wavefunction weight of  the 
outermost node is 
%WJ_18: I tried to distinguish negligible in DFT from zero in TB
% so I added
zero in the tight-binding or
negligible in the DFT approach.
For the same reason, in  a
larger zigzag edge with a single cape structure such as that shown in Fig. \ref{fig4}, the majority of %aay the 
edge atoms 
are from the original zigzag edge, which belong to the same sublattice as the outermost %aay atom 
atoms
of the cape 
structures.
In this case the weight in the
edge state is thus
in the sublattice of the majority of the zigzag edge atoms, with a nonzero value in the outermost cape node, as illustrated 
%LC better add two more
in the right part of  Fig. \ref{fig8}.
%WJ_19: one word added
%right.

\subsection{Mixing and splitting diagrams}

As discussed in the previous Sections, 
the
mixing and splitting of the flat bands wavefunctions occur due to 
the
hybridization of 
orbitals between neighboring nodes, 
which belong to 
different sublattices. 
We introduce simple diagrams %aay, which 
that help to understand %aay the 
such
hybridization. 
We can explain how such bands at 
%WJ_20: math-mode added
$E=0$ 
split and their wavefunctions localize.

Each diagram is built of two rows containing square boxes 
 where each box represents a non-degenerate band at zero energy.
The upper and lower rows correspond to the upper and lower GNR edges, respectively. 
Empty and filled boxes represent bands that localize at different sublattices. %aay Extra node 
Each extra node added to a given edge is  represented by 
a
box added to the corresponding row. The added box is empty or filled depending on the sublattice it belongs to. %aay Now, a 
A pair of empty and filled boxes in a given row  represents 
now 
two
interacting and
 hybridizing 
bands which 
must split and 
move away from zero energy. 
Their
corresponding pair of empty and filled boxes
annihilate
and  disappear 
from the row. The remaining boxes represent the flat bands which survive 
such
hybridization process. Their filling %aay uniquely 
determines the sublattice at which they localize.

We describe next how the diagrams explain the mixing and splitting of the flat bands 
in a practical case. We consider extra nodes
added to the edge of (4,0) GNR to form a ribbon with a
cape 
structure, 
as shown previously in Fig. \ref{fig4}. 

The diagram corresponding to Fig. \ref{fig4}(a), i.e. to the (4,0) zigzag GNR, has for $k < 2/3 \pi$ one filled box in the upper row (representing 
an
 $E=0$ state localized at the upper edge, 
 on the 
 sublattice marked by filled circles) and one empty box in the lower row (representing 
 an $E=0$ state localized at the lower edge,
 on the 
 sublattice marked by empty circles). For $k > 2/3 \pi$ we have two filled boxes in the upper row and two empty boxes in the lower row. We have already described the process of adding two extra Klein nodes %aay and the 
in the previous subsection. The
corresponding diagrams are shown under 
the bands depicted 
in Fig. \ref{fig4}(b) and (c). The two Klein nodes at each side of the GNR unit cell add two empty/filled boxes in the upper/lower rows (Fig. \ref{fig4}(b)). The pairs of empty/filled boxes in each row annihilate (Fig. \ref{fig4}(c)) and the corresponding bands split. %aay Only two 
Two
boxes are 
only
left at the Fermi energy for $k < 2/3 \pi$ as shown in Fig. \ref{fig4}(d). 
 
The (4,0) GNR with a 
cape 
structure has an extra node connected to the existing Klein nodes at each side of the GNR. 
When these extra nodes are not connected, two additional $E=0$ bands appear in the spectrum. They are represented by two extra boxes added to the existing diagram: 
filled box in the upper row, and 
an 
empty box in the lower row.
Now, for $k < 2/3 \pi$ the pair of filled/empty boxes in a given row annihilates, leading to the hybridization and splitting of the zero-energy bands, as shown in Fig. \ref{fig4}(e) and (f). For $k> 2/3 \pi$ the remaining two boxes do not have any partner state to hybridize, %aay so 
therefore
they origin the states at $E=0$ which are localized in the external 
cape 
node and its sublattice. 
Our prescription and diagram analysis confirms the degeneracy and localization 
of bands as seen in the previous Subsection, but without performing any calculations.

\begin{figure}[htbp]
\includegraphics*[width= 80mm]{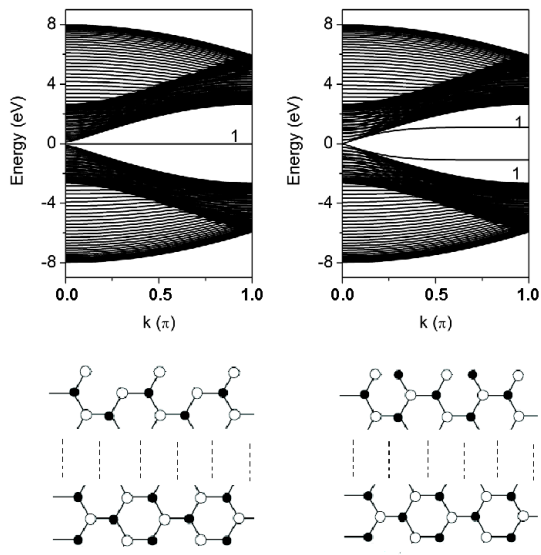}
\caption{Spectra of armchair GNRs with Klein nodes added to the edge. Left: 40(1,1) armchair GNR with one Klein node per unit 
cell. 
Right: the same GNR but with two Klein nodes per unit cell.} \label{fig_arm}
\end{figure}

\subsection{Gap states away from the Fermi energy}

We have already observed (see Fig. \ref{fig4}(d)) that in some cases the split bands do not reach 
the
bulk continuum and for some range of $k$ they form gap states with $E \ne 0$. Similar bands occur in the 
bearded edges investigated in Ref. \onlinecite{wakabayashi_2010}, where zigzag 
edges and Klein 
defects appear alternatively.
\cite{nota_b} Gap states with $E \ne 0$ 
are usually related to 
 edges of mixed character. In this section we consider only two 
 examples 
 of such edges and explain their origin in more detail 
 as an illustration of how 
 our prescription applies to any kind of edges of mixed character.

\subsubsection{Klein defects in armchair 
edges}

Let us consider the case of 
an
armchair ribbon with one edge modified by attaching a Klein node.  The corresponding spectrum is shown in
the upper left panel of Fig. \ref{fig_arm}.
When both edges are modified the flat band is 
doubly 
degenerate.\cite{wakabayashi_2010,note2} It is 
noteworthy
that the same flat band appears 
no matter 
whether the Klein node is disconnected or attached.
Our prescription explains 
this fact 
in a simple way:
such a flat band has no partner band to hybridize and split. 
When a second Klein node is attached (see the bottom right panel of Fig. \ref{fig_arm}), 
the flat bands mix and split
because they belong to different sublattices. 
However, they do not reach 
the band continuum and become $E \ne 0$ gap states.
When we additionally connect 
the previous
two Klein nodes %aay from the neighbor unit cells 
we obtain again  
an armchair GNR. In this case the bands split 
even
farther and merge into the continuum; 
we recover the spectrum of armchair GNR.

\subsubsection{Chiral edges
}

In this Subsection we present the results 
corresponding to general $(n,m)$ edges, i.e., those which are not purely 
armchair or zigzag, either minimal or modified. 
Their borders with Klein defects 
and the corresponding spectra 
can be explained by the folding rule and diagrams 
presented 
above.

As example, we investigate a 3(8,1) GNR with 
a
minimal edge and two different modifications. Studying different edges for a given translation vector ${\bf T}$ 
is important, since different edge modifications are required sometimes to form junctions between graphene edges,
as it happens when 
constructing junctions between carbon nanotubes.
Such studies suggest whether localized or resonance interface states appear at the junctions and even allow to estimate their energies\cite{santos}. 

\begin{figure}[htbp]
\includegraphics*[width= 80mm]{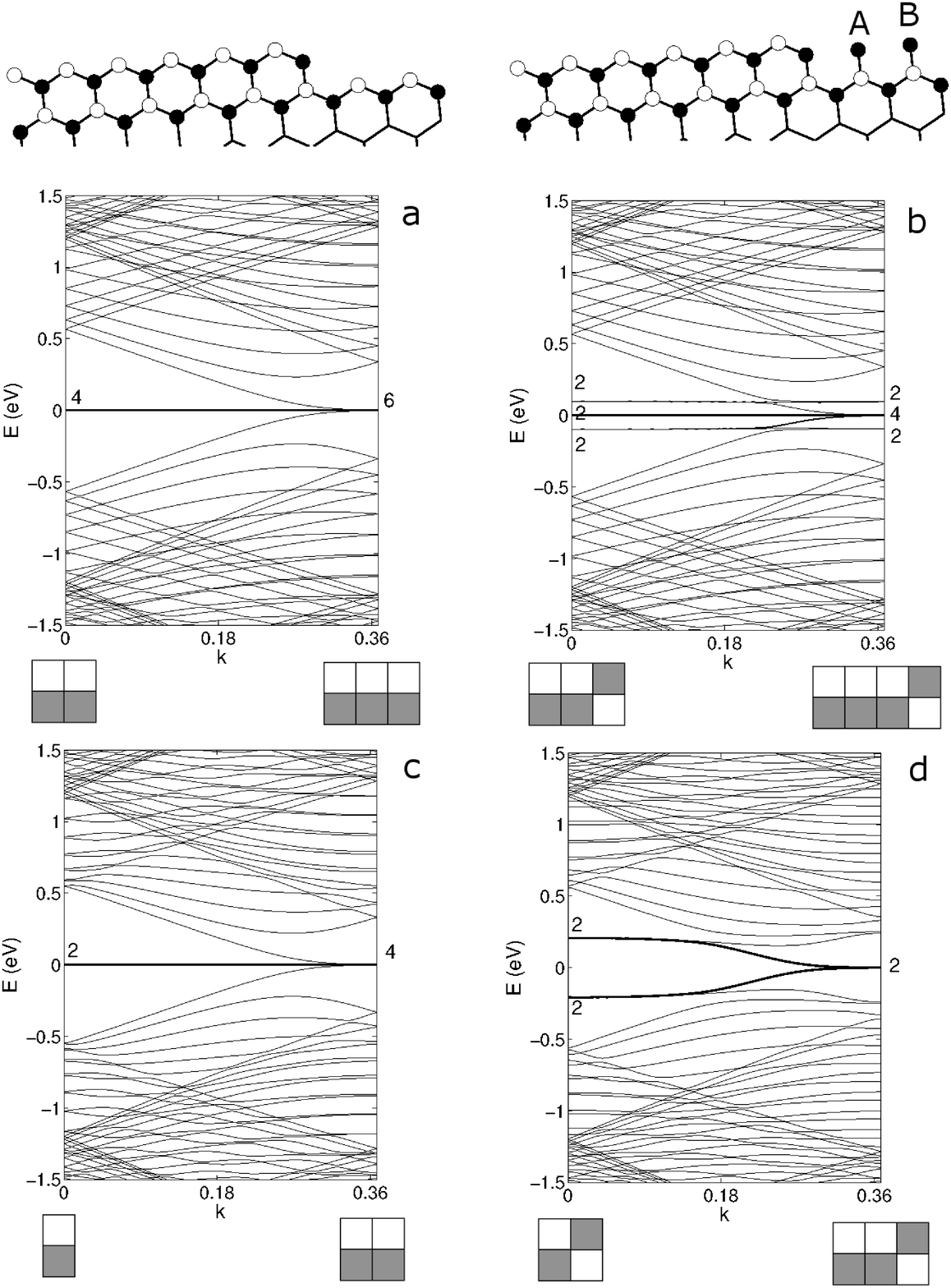}
\caption{
Spectra of 3(8,1) GNR close to the Fermi energy with different edge terminations, ranging from the minimal edge (a) to the modified edge with one Klein node, marked as A, added (b-c) or two Klein nodes A and B attached (d).  In (b) $t=0.1 t_0$, while in (c) and (d) $t=t_0$.  %aay Top panel shows 
The top panels show 
the minimal and modified edges.}\label{fig5}
\end{figure}

The top panel of Fig. \ref{fig5} shows the different edges investigated: minimal (left) and modified (right) by attaching one, A,  or two, A and B, Klein nodes. 
The spectrum of the GNR with minimal edges close to $E_F$ is shown in Fig. \ref{fig5}(a). Note that it is the same spectrum as in Fig. \ref{rib70_81}(right). The degeneracy of the $E=0$ band results from the folding rules corresponding to the (7,0) GNR. When an extra 
non-connected node (marked as A in the Figure) is added at each side of the GNR unit cell, 
a doubly degenerated $E=0$ band
appears 
 in the spectrum. 
Thus, the degeneracies 
 increase to six and eight for $k< 2/3 \pi$ and $k> 2/3 \pi$, respectively. 
When we connect this extra node by $t=0.1 t_0$, 
two 
flat bands hybridize and split, see Fig. \ref{fig5}(b).
The hybridization and splitting can be explained by the attached diagrams. 
For $t=t_0$ 
the split bands merge into the ribbon continuum of states and disappear from the gap region; the spectrum and the corresponding diagrams are shown in Fig. \ref{fig5}(c).
If the other Klein node marked B is added to this edge, first as a non-connected node, the 
degeneracy of the $E=0$ ribbon band increases by two. If the connection of the atom B 
is turned on, the bands mix and split.  
Fig. \ref{fig5}(d) shows the bands for $t= t_0$. All the boxes in the diagram corresponding to $k < 2/3 \pi$ 
annihilate. 
Only a
doubly degenerate zero-energy band survives %aay , extending 
for $k > 2/3 \pi$. In the corresponding diagram two pairs of empty and filled boxes 
annihilate, leaving only one box in each row %aay, what exactly 
that
corresponds to a
doubly degenerate %aay zero-energy 
band at zero energy. 
The attached diagrams
also indicate that 
the flat bands 
in both cases (c) and (d) 
localize in the sublattices 
corresponding to the zigzag edges. %aay Again, numerical calculations fully confirm these predictions.
Our numerical calculations fully confirm again these predictions.

A comment is required to explain why the addition of the second Klein node B yields weaker splitting of the flat bands than in the case where only a node A is attached. A closer inspection of the two wavefunctions localized at one edge at $k=0$ in Fig. \ref{fig5}a reveals that one of them is %aay closer 
nearest
to a step in the edge, see upper panel. %aay Thus, such 
Such
step-localized wavefunction is
thus
nearer to the Klein defects. The second wavefunction
at B
localizes away from the step. Attaching the first Klein node A yields a strong hybridization with the step-localized wavefunction. The
wavefunction of the 
second Klein node
B
must hybridize with the 
remaining function, 
which localizes away from the Klein node. This explains why the %aay second 
mixing
of B
is weaker and %aay the 
its
splitting smaller.

\section{Summary}
%WJ_20: I have changed to "present perfect"
% and added a sentence about the connection of flat band
% splitting with the GNR width
\label{sec:sum}
We have presented
a simple prescription %aay allowing 
to predict, without performing any calculations, the existence of zero-energy bands in graphene nanoribbons 
and graphene edges of arbitrary shape.
Our prescription is based on two observations: (a) any edge can be created from a 
 minimal edge by adding extra nodes, (b) the zero-energy spectra of 
 graphene minimal edges 
 are uniquely defined by the edge band structure of its 
 zigzag $(n,0)$ 
 component, which in turn is obtained by 
 folding $n$ times the spectrum of the 
 zigzag (1,0) edge.
  Extra but disconnected nodes provide zero-energy states %aay, 
which, after connecting them to the graphene edge,
  may 
 hybridize with the existing flat bands and split. 
The splitting occurs only when the extra node belongs
to a different sublattice as that where the edge zero-energy
bands are localized.
We have 
introduced simple rules and diagrams allowing to precisely determine not only the existence of the flat bands, but also their degeneracies and localization of their wavefunctions on the graphene sublattices. The folding rules allow also to estimate the energy gaps
which open in certain regions of $k$.
%WJ_21: two sentences added - 
Our prescription and diagrams hold for graphene edges and nanoribbons with arbitrary geomeries. However, one has to remember that for extremely narrow ribbons the interaction between the GNR edges may lead to the edge bands splitting and the loss of their flatness.

We have 
considered a number of GNRs with different edges, some of them with attached Klein nodes  
or cape structures. 
They are also important 
in the study of graphene-based 
%WJ_21: one word added
complex
systems, since connecting 
different portions of graphene requires sometimes 
the modification of their edges. 

Finally, we have shown 
that our prescription %aay correctly 
predicts the localization properties of edge states. We have 
 studied %aay it for 
a couple of cases by performing tight-binding and first-principles DFT calculations. %aay, although 
The correspondence between the ab-initio and the tight-binding results back our prescriptions, 
with the caveat that within the DFT approach the edge bands are no longer exactly flat and with zero-energy. 
Our method can be widely used to foresee the existence of edge states and flat bands in any graphene edges and ribbons.

\section{Acknowledgments}

We acknowledge financial support from Basque Departamento de Educaci\'on and the
UPV/EHU (Grant No. IT-366-07), the Spanish Ministerio de Innovaci\'on,
Ciencia y Tecnolog\'{\i}a
(Grants  No. FIS2007-66711-C02-02 and FIS2009-08744)  and  the ETORTEK  research
program funded  by  the  Basque
Departamento de Industria and the Diputaci\'on Foral de Guip\'uzcoa.
W.J. thanks Donostia International Physics Center,  and A.A. and L.C. thank Nicolaus Copernicus University in Toru\'n for hospitality. 


\begin{thebibliography}{0}
\bibitem{note2} When both Klein nodes belong to the same sublattice the degeneracy is exact for all $k$, otherwise the bands will split for narrow ribbons.
\bibitem{nota_b} The band structure of 
%LC
these ribbons with bearded edges can be easily explained with our prescription. It is sufficient to consider a (2,0) GNR with only one Klein defect added at each side of the GNR unit cell.
\end{thebibliography}
\end{document}